\documentclass[conference]{IEEEtran}
\IEEEoverridecommandlockouts

\usepackage{cite}
\usepackage{amsmath,amssymb,amsfonts}
\usepackage{algorithmic}
\usepackage{graphicx}
\usepackage{textcomp}
\usepackage{xcolor}
\usepackage{mathtools}
\usepackage{booktabs}

\usepackage{glossaries}

\newacronym{iot}{IoT}{Internet of Things}
\newacronym{rle}{RLE}{run-length encoding}
\newacronym{quars}{QuaRs}{Quantile Reshuffling}
\newacronym{aad}{AAD}{average absolute difference}
\newacronym{lz}{LZ}{Lempel-Ziv}
\newacronym{lzss}{LZSS}{Lempel-Ziv Storer-Szymanski}
\newacronym{ans}{ANS}{Asymmetric Numeral System}
\newacronym{fse}{FSE}{Finite State Entropy}
\newacronym{zstd}{Zstd}{Zstandard}
\newacronym{iid}{i.i.d.}{independent and identically distributed}

\def\BibTeX{{\rm B\kern-.05em{\sc i\kern-.025em b}\kern-.08em
    T\kern-.1667em\lower.7ex\hbox{E}\kern-.125emX}}

\begin{document}

\title{Lossless Compression of Time Series Data:\\ A Comparative Study
}

\author{\IEEEauthorblockN{Jonas G. Matt}
\IEEEauthorblockA{\textit{Automatic Control Laboratory} \\
\textit{ETH Z\"urich}\\
Z\"urich, Switzerland \\
jmatt@ethz.ch}
\and
\IEEEauthorblockN{Pengcheng Huang}
\IEEEauthorblockA{\textit{Corporate Research Center} \\
\textit{ABB}\\
Baden-D\"attwil, Switzerland \\
pengcheng.huang@ch.abb.com}
\and
\IEEEauthorblockN{Balz Maag}
\IEEEauthorblockA{\textit{Corporate Research Center} \\
\textit{ABB}\\
Baden-D\"attwil, Switzerland \\
balz.maag@ch.abb.com}
}

\maketitle

\begin{abstract}
    Our increasingly digital and connected world has led to the generation of unprecedented amounts of data.
    This data must be efficiently managed, transmitted, and stored to preserve resources and allow scalability. 
    Data compression has therein been a key technology for a long time, resulting in a vast landscape of available techniques.
    This largest-to-date study analyzes and compares various lossless data compression methods for time series data.
    We present a unified framework encompassing two stages: data transformation and entropy encoding.
    We evaluate compression algorithms across both synthetic and real-world datasets with varying characteristics.
    Through ablation studies at each compression stage, we isolate the impact of individual components on overall compression performance -- revealing the strengths and weaknesses of different algorithms when facing diverse time series properties.
    Our study underscores the importance of well-configured and complete compression pipelines beyond individual components or algorithms; it offers a comprehensive guide for selecting and composing the most appropriate compression algorithms tailored to specific datasets.
\end{abstract}


\section{Introduction}

The proliferation of time series data in today's digital landscape reaches unprecedented levels.
For the industrial sector, the rapid emergence of Industry 4.0 has catalyzed \gls{iot} technologies, resulting in an exponential increase in data generation rates.
Estimates from recent studies suggest that in 2025, \gls{iot} devices alone will generate more than 180 zettabytes ($10^{21}$ bytes) of new data per year \cite{DataGrowthWorldwide}.
Reducing the amount of this data is paramount to maximize the efficiency of its transmission, processing, and storage.\par
One of the most general and effective approaches for reducing data volume is lossless data compression.
Lossless compression exploits statistical redundancy in the data to represent it more efficiently but at no loss of information.
It stands out from other data reduction techniques, such as lossy compression, decreasing sampling frequency or numerical accuracy, because it can be applied independently without compromising the data being collected.\par
State-of-the-art lossless compression techniques were designed primarily for text or unstructured binary data rather than numeric datasets~\cite{deutschDEFLATECompressedData1996,sewardBzip22019,colletZstandardCompressionApplication2021,GoogleBrotli2024}.
In fact, the most relevant compression benchmarks used today focus almost exclusively on text data \cite{mahoneyLargeTextCompression2011,InikepLzbench}, leaving out important aspects related specifically to compressing numeric sequences.
This has sparked the proposal of new methods in recent years~\cite{mogahedDevelopmentLosslessData2018,spiegelComparativeExperimentalStudy2018a,blalockSprintzTimeSeries2018,vestergaardTitchyOnlineTimeSeries2021,loncaricPcodecLosslessCodec2023}.
Yet, in the authors' opinion, compression techniques tailored to such data remain understudied.
Existing works on the compression of time series either appear as pure surveys of literature~ \cite{chiarotTimeSeriesCompression2023, deoliveiraTimeSeriesCompression2023}, or they propose new methods including comparisons against other baseline algorithms \cite{mogahedDevelopmentLosslessData2018,spiegelComparativeExperimentalStudy2018a,blalockSprintzTimeSeries2018,goyalDZipImprovedGeneralpurpose2020,vestergaardTitchyOnlineTimeSeries2021,loncaricPcodecLosslessCodec2023}.
Neutral comparisons the authors are aware of are limited in the number and variety of algorithms they feature \cite{bjarasComparativeStudyCompression2019, hughesComparisonLossyLossless2023} and/or were designed with specific applications in mind \cite{ketshabetsweDataCompressionAlgorithms2021,pasettiComprehensiveEvaluationLossless2023,bozhilovSystematicSurveyCompression2024}.\par
Hence, there is a clear need for an extensive systematic comparison of existing methods to provide guidance for choosing appropriate algorithms.
This paper presents the largest-to-date comparison of lossless compression algorithms applicable to time series.
We include both general-purpose compression algorithms as well as new methods targeted at time series.
We provide a common framework encompassing those techniques and identify the key steps involved within time series compression pipelines.
By ablating the most common building blocks of such pipelines, we highlight the impact of various components given different signal characteristics.
A special emphasis is put on evaluating the effects of compression-aiding data transformations such as delta coding, which are employed before the actual data compression.
By granting equal access to these transformations, fair assessments can be performed to conclude the impact of various methods adopted in different stages of the compression pipelines, which are often underrepresented in the literature.
Overall, our research highlights the importance of well-configured and holistic compression pipelines, surpassing the focus on individual components or algorithms. Our study provides a comprehensive framework and guidance for selecting and integrating the most suitable compression algorithms for specific time series datasets.\par
The rest of this paper is organized as follows: In Section \ref{sec:definitions-and-methods}, the necessary notation and metrics are defined, and the compared compression methods are introduced.
Section \ref{sec:framework} describes the proposed framework for compression pipelines.
In Section \ref{sec:experiments}, we present the results of the comparative study, and we conclude and summarize the paper in Section \ref{sec:conclusion}.
\section{Definitions and methods}
\label{sec:definitions-and-methods}

\subsection{Time Series}
\label{subsec:time_series}

An (integer-valued) time series $\mathcal{T}$ is a chronologically ordered sequence of $n$ samples $\mathcal{T} \coloneq (x_k)_{k=1}^n = (\mathbf{x}_1, ...
, \mathbf{x}_n)$ obtained from the same source, where $\mathbf{x} \in \mathbb{Z}^d$.
If $d = 1$, the time series is called univariate, and multivariate otherwise.
Note that industrial time series from sensor readings can often be ``integerized'', as the original readings are performed by analog-to-digital converters with fixed numbers of total values.
This is why we focus on integer time series in this study.

\setcounter{subsubsection}{0}
\subsubsection{Delta of a time series}
The delta of a time series yields another time series that contains the differences between consecutive samples of the original time series.
Formally, the delta of a time series $(x_k)_{k=1}^n$ is defined as $(\Delta x_k)_{k=1}^n \coloneq (x_1, x_2-x_1, \dots, x_n-x_{n-1})$.
One can view delta coding as a common decorrelation technique for time series data, allowing to transmit only new information or relative entropy.

\subsubsection{Average absolute deviation and cardinality}
We define two distinct measures for the ``spread'' of a time series: its \gls{aad} and its cardinality.
The \gls{aad} is the average of the absolute deviations of a time series' samples from some center value.
For our purpose, we define the center value as $0$.
The cardinality is the number of distinct values in the time series; it serves as a measure of a signal's information or entropy.

\subsubsection{Compression ratio and score}
Arguably, the most important aspect of any compression method is its ability to shrink the size of the data.
A straightforward way to measure this is the compression ratio $\text{CR} \coloneq \frac{\text{original size}}{\text{size after compression}}$,
with the size of both the source and encoded data measured in bytes.
By definition, the compression ratio is unbounded.
To facilitate the direct comparison of methods in a more intuitive manner, we define the compression score as $\text{CS} \coloneq 1 - \frac{1}{\text{CR}}$.
The compression score can be interpreted as the fraction of the file size removed by compression.

\subsubsection{Compression time and speed}
Comparing the computational complexity of different compression algorithms experimentally is more difficult than comparing their compression abilities.
A fair comparison is impeded by differences in implementation and other external factors.
Nevertheless, a common approximation of a compression method's time complexity is compression time.
It is defined as the computation time in seconds that a given algorithm requires to compress a given data.
Further, we define the compression speed as the megabytes of data compressed in one second, with unit MB/s.
We refrain from making any statements about memory requirements of the reviewed methods because we cannot guarantee a fair comparison.

\subsection{Compression Algorithms}

We briefly review state-of-the-art lossless compression algorithms applicable to time series.
We provide a unified framework in Section~\ref{sec:framework} to compare those algorithms.
All reviewed algorithms are open source.

\textit{\underline{Entropy coding:}}
As a fundamental technique in compression, entropy coding leverages the statistical properties of data to assign shorter codes to more frequent symbols, thereby reducing the overall data size.
It aims to approach the theoretical limit of compression efficiency as defined by Shannon's source coding theorem~\cite{shannonMathematicalTheoryCommunication1948}.

\setcounter{subsubsection}{0}
\subsubsection{Huffman coding~\cite{huffmanMethodConstructionMinimumRedundancy1952}}

This is an optimal prefix coding algorithm widely used in lossless data compression.
It begins by creating a priority queue of nodes, each representing a symbol and its frequency.
The two nodes with the lowest frequencies are repeatedly merged to form a binary tree.
The path from the root to each leaf node determines the symbol's binary code, with shorter codes for more frequent symbols.

\subsubsection{Arithmetic coding~\cite{saidIntroductionArithmeticCoding2023}}
Instead of assigning fixed codes to individual symbols, arithmetic coding represents the entire message as a single number between 0 and 1.
Starting with the interval [0,1), each arriving symbol refines the interval based on its probability, assigning larger subintervals to more frequent symbols.
This process continues iteratively until the final interval encodes the entire message.

\subsubsection{Asymmetric numeral systems~\cite{dudaAsymmetricNumeralSystems2009}}
First proposed by Jaros\l{}aw Duda in 2009, \gls{ans} are a modern family of entropy coders.
\gls{ans} guarantee compression close to the Shannon limit for all distributions, all while having a time complexity similar to Huffman coding.

\textit{\underline{Dictionary coding: }}
This category of classical compression algorithms works by representing repetitive patterns in data with dictionaries.

\setcounter{subsubsection}{0}
\subsubsection{LZSS~\cite{storerDataCompressionTextual1982}}
\gls{lzss} is one of the most popular variants of the \gls{lz} algorithm family.
It improves on the basic LZ77~\cite{zivUniversalAlgorithmSequential1977} and LZ78~\cite{zivCompressionIndividualSequences1978} algorithms by omitting dictionary references that are longer than the sequence they would replace.
Like most general-purpose algorithms, \gls{lzss} acts on the source data in a byte-wise manner and is agnostic to the underlying data type.

\subsubsection{LZ4~\cite{colletLZ4ExtremelyFast2014}}
As a modern algorithm based on \gls{lz} dictionary coding, LZ4 focuses on compression speed.
By utilizing the capabilities of modern CPUs, it can achieve more than 500 MB/s per core during compression and multiple GB/s per core during decompression.

\subsubsection{Snappy~\cite{SnappyFastCompressor2014}}
Snappy was developed by Google as another algorithm for high-speed requirements.
Like LZ4, it is a modern reinterpretation of the \gls{lz} algorithms.

\textit{\underline{Hybrid general-purpose methods:}}
This class of methods combines multiple stages to achieve better compression, following the framework discussed in Section~\ref{sec:framework}.

\setcounter{subsubsection}{0}
\subsubsection{zlib~\cite{deutschZLIBCompressedData1996}}
One of the most popular hybrid methods is the Deflate algorithm \cite{deutschDEFLATECompressedData1996}.
It consists of a dictionary coding stage based on LZ77 followed by Huffman coding.
Deflate is among the most widely used compression algorithms and is part of popular archivers such as ZIP and 7-Zip;
zlib is a popular implementation of Deflate.

\subsubsection{LZMA~\cite{LZMASDKSoftware}}
LZMA combines LZ77-like dictionary coding with range coding, an implementation of arithmetic coding.
Owed to its arithmetic coding stage, LZMA requires more memory and compresses more slowly than Deflate.

\subsubsection{Brotli~\cite{GoogleBrotli2024}}
Brotli is a general-purpose compression algorithm developed by Google and released in 2013.
It combines a modern version of LZ77 dictionary coding with Huffman entropy coding and second-order context modeling.
Various optimizations enable it to achieve better compression ratios than Deflate, at similar speeds and memory requirements.

\subsubsection{Zstandard~\cite{colletZstandardCompressionApplication2021}}
With the emergence of \gls{ans} as a new type of entropy coder, their usage in hybrid methods was explored by, e.g., Facebook and gave rise to \gls{zstd}.
It is based on Yann Collet's \gls{fse} \cite{colletFiniteStateEntropy2013}, a tabled implementation of \gls{ans}.
Overall, \gls{zstd} combines an LZ77 dictionary coder with a two-stage entropy coder featuring both Huffman coding and \gls{fse}.

\subsubsection{bzip2~\cite{sewardBzip22019}}
In contrast to the other hybrid methods, bzip2 does not feature a dictionary coder.
Instead, it applies a sequence of \gls{rle}~\cite{robinsonResultsPrototypeTelevision1967}, a Burrow-Wheeler transform~\cite{burrowsBlocksortingLosslessData1994} (BWT), a move-to-front transform (MTF) ~\cite{bentleyLocallyAdaptiveData1986}, and another \gls{rle}.
The output of the final \gls{rle} is Huffman-coded.

\subsubsection{Blosc~\cite{altedBloscBlockingShuffling2018}}
Blosc is a lossless compression algorithm primarily designed for scientific computing and large datasets.
It employs a blocking strategy to split data into chunks that are independently shuffled and compressed.
Using multi-threading, Blosc achieves low latency and high throughput.

\textit{\underline{Integer compression:}}
Specialized methods for integer compression exploit the separation of numbers at their actual byte boundaries (e.g., taking two bytes at a time for 16-bit integers).
Then, they aim to store each number as efficiently as possible given their distribution.
Most methods make an implicit assumption about the numerical distribution of the data.
Most commonly, smaller numbers are assumed to occur more frequently than larger numbers.
Thus, compression is achieved by assigning shorter codes to smaller numbers.
We refer to this as the ``smaller-numbers-fewer-bits'' principle.

\setcounter{subsubsection}{0}


\subsubsection{Exp-Golomb~\cite{teuholaCompressionMethodClustered1978}}
Exponential Golomb is a universal code that combines unary and binary coding; it represents a number $n$ as a sequence of zeros followed by the binary representation of $n+1$.
The length of the zero sequence is one less than the length of the following binary representation.

\subsubsection{Bit packing}
Bit packing aims at storing integers at the minimum number of bits required.
The number of bits is determined by the size of the largest integer in a block of data.
For each block, the number of bits is stored once, followed by the integers in that block.
An example of bit packing algorithms is the Simple family~\cite{trotmanOptimalPackingSimpleFamily2015}.

\textit{\underline{Time series compression:}}
Recently, a few specialized methods have been proposed for the compression of time series.

\setcounter{subsubsection}{0}
\subsubsection{DRH~\cite{mogahedDevelopmentLosslessData2018}}
DRH proposes to compress time series via the three following stages: delta coding, \gls{rle}, and a static Huffman code that is inspired by the JPEG standard \cite{pennebakerJPEGStillImage2004}.
Delta coding and run-length encoding make up the decorrelation stage and the static Huffman code is the entropy coding stage.
We use a custom implementation of the static Huffman code.

\subsubsection{Sprintz~\cite{blalockSprintzTimeSeries2018}}
Sprintz is a time-series-specialized compression algorithm that employs a combination of delta coding, run-length encoding, and bit-packing.
A forecasting-based encoding strategy called FIRE (Fast Integer REgression) may be used in place of delta coding.
Since the improvements of FIRE are small (according to the authors of Sprintz and our own experiments), we omit this variant of Sprintz.
Focussing on delta coding also fits the ablation study of this paper.

\subsubsection{Pcodec~\cite{loncaricPcodecBetterCompression2025}}
Pcodec is an open-source compression method designed for numerical sequences.
It uses a three-step procedure: (1) mode decomposition to split numbers into latent variables based on their approximate structure, (2) delta coding to compress differences between consecutive elements, and (3) quantile-based binning, its core innovation, to entropy-code latent variables into bins paired with exact offsets.
Its binning strategy enables Pcodec to approach the Shannon limit for any smooth numerical distribution.

\subsection{Methods considered but not included}

The following methods were considered but did not find their way into the experiments of this paper.

\subsubsection{FastPFor~\cite{lemireDecodingBillionsIntegers2015}}
A library for fast integer compression that supports a variety of different codecs.
However, it is limited to 32-bit integers.

\subsubsection{Titchy~\cite{vestergaardTitchyOnlineTimeSeries2021}}
A recently proposed time series compression algorithm that is specifically designed for random access but is without an official implementation.

\subsubsection{Gorilla~\cite{pelkonenGorillaFastScalable2015}}
A time series compression algorithm designed for floating point data.
Due to the generally lower compressibility of floats, it is not competitive on integer data.

\subsubsection{Other universal codes}
We omit other static integer codes such as Rice~\cite{riceAdaptiveVariableLengthCoding1971} or Fibonacci~\cite{apostolicoRobustTransmissionUnbounded1987} codes as they would not yield additional insights next to Exp-Golomb and DRH.
\section{A unified framework for compression pipelines}
\label{sec:framework}

We have seen in Section~\ref{sec:definitions-and-methods} that there is a plethora of lossless compression algorithms applicable to time series data, all following different design principles and even encompassing different modules/components within a single compression algorithm.
A unified framework for compression is therefore needed to \textit{(i)} identify the common components of compression algorithms, \textit{(ii)} perform ablation study of such components and study individually their impacts on overall compression, \textit{(iii)} fairly compare different compression algorithms by giving them access to common fundamental components and \textit{(iv)} guide the composition of new compression pipelines by combining different building blocks within this unified framework.\par
Our proposed unified compression framework consists of two stages, compression-aiding transformations, and entropy coding.
For each stage, many different algorithms exist following possibly drastically different design philosophies.\par
Entropy coding aims at finding the most efficient bit representations (codes) for the source symbols, achieving compression close to the data's entropy.
Entropy coders represent the source symbols independently and implicitly assume that they are \gls{iid}.
However, real data often has internal correlations, leading to redundancy.
Reducing such redundancies, or entropy of the data in more theoretical terms, is thus helpful in enhancing the performance of entropy coders.
Compression-aiding transformations aim to reduce the amount of information to be encoded or to prepare the data such that it can be better encoded by certain entropy coders.\par
\begin{figure}
	\centering
	\vspace{.2em}
	\includegraphics[width=\columnwidth,trim={0 0 0 0},clip]{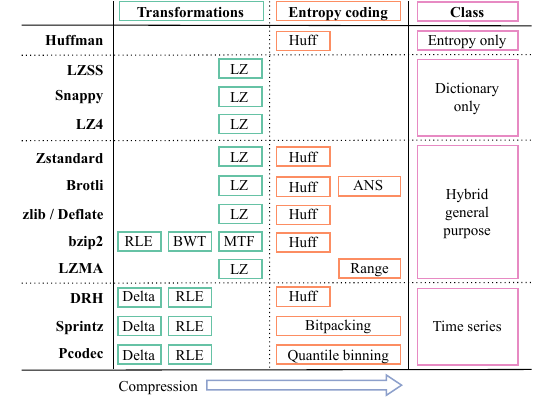}
	\vspace{-1em}
	\caption{Overview of the main compared methods categorized by the unified compression framework, detailing their transformation and entropy coding stages.}
	\label{fig:methods}
    \vspace{-1em}
\end{figure}

Figure~\ref{fig:methods} presents a summary of selected representative compression algorithms as discussed in Section~\ref{sec:definitions-and-methods}, according to our compression framework.
The two-stage concept is the foundation of existing hybrid methods~\cite{deutschDEFLATECompressedData1996,deutschZLIBCompressedData1996,GoogleBrotli2024,colletZstandardCompressionApplication2021} as well as methods tailored to time series data~\cite{blalockSprintzTimeSeries2018,loncaricPcodecLosslessCodec2023,mogahedDevelopmentLosslessData2018}.\par
With Figure~\ref{fig:methods}, we provide a more complete picture of how new tailored compression methods could be constructed for specific use cases.
In the following, we discuss the two stages and some common design principles employed.

\subsection{Compression-aiding transformations}
\label{subsec:compression-aiding-transforms}

The following transformations (among others) are applied before entropy coding to convert internal correlations into higher compression ratios.

\subsubsection{Dictionary coding}
For many types of data, the correlation between source symbols manifests itself in the recurrence of identical sequences of symbols. 
Occurrences of the same word in a text are just one example, occurrences of identical values in a time series are another.
Dictionary coding techniques enhance compression performance by removing such redundancy.
They constructs a dictionary of recurring symbol sequences and replace source sequences with their dictionary positions.
Implementations such as LZ77 and its variants (LZ78, \gls{lzss}, LZW)~\cite{zivUniversalAlgorithmSequential1977,zivCompressionIndividualSequences1978,storerDataCompressionTextual1982,welchTechniqueHighPerformanceData1984} use a sliding window to store substrings of processed data, limiting memory usage.
Matches in the sliding window are replaced by their relative positions and lengths.

\subsubsection{Delta coding}
Not without reason, delta coding is part of virtually all compression methods that specialize in time series~\cite{mogahedDevelopmentLosslessData2018,blalockSprintzTimeSeries2018,loncaricPcodecLosslessCodec2023}.
It decorrelates data by transforming it into a series of differences between consecutive values, rather than the values themselves.
This process reduces the correlation between data points, making the data more uniform and predictable.
By encoding the differences (deltas) instead of the actual values, delta coding captures the small variations between data points, reducing redundancy.

\subsubsection{Run-length encoding~\cite{robinsonResultsPrototypeTelevision1967}}
A feature commonly found in real data is consecutive repetitions of the same symbol.
In a time series, this corresponds to a signal that remains constant for some time.
\gls{rle} replaces such ``runs'' by a pair consisting of the symbol itself and the number of repetitions.
Given some prior knowledge of the data, it can be beneficial to apply \gls{rle} only to a subset of source symbols.
For instance, in the case of time series deltas, it makes sense to apply \gls{rle} only to runs of zeroes.
This is the variant of \gls{rle} used in our experiments.

\subsubsection{Quantile reshuffling~\cite{mattQuaRsTransformBetter2025}}
\gls{quars} aims at facilitating compression techniques that encode smaller numbers (smaller \gls{aad}) with fewer bits.
It reshapes arbitrary numerical data distributions into unimodal distributions centered around zero; during the reshuffling, more frequent numbers (ordered by quantile bins) are mapped to smaller values.
This way, it effectively reduces the \gls{aad} of a dataset.

\subsection{Entropy coding}
\label{subsec:entropy_coding}
Once data is preprocessed with potential entropy-reducing and compression-aiding transforms, the entropy coding stage then tries to find the most efficient codes or bit representations for the symbols in the transformed data.
It was the traditional focus of compression research and covers a wide variety of algorithms.
Existing methods can be categorized as follows.

\subsubsection{Dynamic codes}
These codes are designed based on a given data distribution.
Some information on the data's statistical properties is also required for decoding, and must hence be transmitted along with the compressed data. 
Concrete methods include Huffman coding, arithmetic coding, and \gls{ans}. 

\subsubsection{Universal codes}
In contrast, universal codes are designed to efficiently encode integers without prior knowledge of data distribution.
Typically, they assign shorter codes to smaller integers, following the smaller-numbers-fewer-bits principle.
Common methods include Fibonacci coding~\cite{apostolicoRobustTransmissionUnbounded1987}, Exp-Golomb coding, and Elias codes~\cite{eliasUniversalCodewordSets1975}.
Each method can be associated with an induced probability distribution, for which it represents an optimal prefix code.


\section{Experiments}
\label{sec:experiments}

The following experiments evaluate the algorithms described in Section \ref{sec:framework} on various types of time series data.
In the following Section (\ref{subsec:experimental-setup}), we state the specifications of the experimental setup.
In Section \ref{subsec:test_case_suite}, synthetic data is used to demonstrate the isolated effect of different signal characteristics on compression performance.
Section \ref{subsec:real-world-time-series-data} presents the results of a comparison on real-world datasets.

\subsection{Experimental setup}
\label{subsec:experimental-setup}

The experiments are conducted on a Linux machine with an Intel Xeon Silver 4208 CPU (2.10 GHz) and 96 GB of RAM.
Unless stated otherwise, we use the methods at their highest or close-to-highest settings: Compression level 9 for zlib, level 19 for Zstd, level 10 for Brotli, level 9 for bzip2, level 6 for LZMA, level 9 and BloscLZ as the dictionary coder for Blosc.
We use delta coding as the forecaster for Sprintz, and level 12 and no higher-order delta coding for Pcodec.
For LZSS, we use a buffer size and lookahead buffer size of 4096 and 18 bytes, respectively.

\subsection{Synthetic time series data}
\label{subsec:test_case_suite}

We start by investigating the impact of isolated signal characteristics on compression performance.
We focus on a comparison of compression scores in this section and defer the discussion of compression speeds to Section \ref{subsec:real-world-time-series-data}. In both sections, we number our main findings to enhance readability.\par
Real data can be interpreted as a superposition of various basic signal characteristics.
With this in mind, we use a suite of synthetic time series to represent a selection of these basic characteristics.
Thus, each test case represents one particular signal characteristic that may occur in time series data.
The selected signals are as follows:
\subsubsection{Sine}
An example of a signal that varies slowly relative to the sampling frequency.
This is where delta coding is the most effective.
\subsubsection{Noise}
Uniformly and independently distributed random noise.
Such data has the maximum theoretical entropy, making it the worst case for compression.
\subsubsection{Sine + noise}
The sum of the previous two signals.
This case illustrates noise can be a bottleneck for compression.
\subsubsection{Switching}
A signal switching between a discrete set of values at short intervals.
This yields a sparse signal with low cardinality but high \gls{aad}.
It is a realistic failure case for algorithms that rely on the efficient encoding of small integers.

\begin{figure}[ht]
    \centering
    \includegraphics[width=\columnwidth,trim={.1cm 0 .5cm 0},clip]{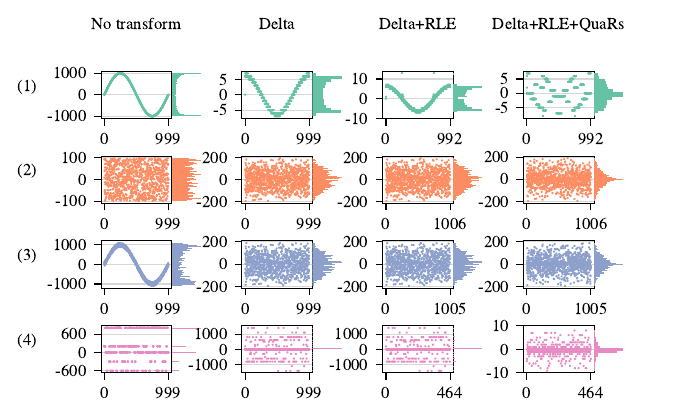}
    \vspace{-2em}
    \caption{
        The synthetic test cases, before and after progressively applying the different transformations (delta coding, \gls{rle}, \gls{quars}).
        This demonstrates the impact of each transformation stage on data representation.
    }
    \label{fig:test_cases-a}
    \vspace{-1em}
\end{figure}

Fig. \ref{fig:test_cases-a} depicts the synthetic data along with its transformed variants (delta coding, \gls{quars}).
Table \ref{tab:cardinality-aad} lists their cardinality and \gls{aad}, before and after applying the transformations.
\begin{table}[t]
    \centering
    \caption{Cardinality / \gls{aad} of the synthetic time series before and after progressively applying the transformations. D, R, and Q denote delta coding, \gls{rle}, and \gls{quars}, respectively.}
    \begin{tabular}{lllll}
\toprule
 & No transform & D & D+R & D+R+Q \\
\midrule
Sine & 897 / 635.5 & 15 / 4.0 & 16 / 4.1 & 16 / 2.7 \\
Noise & 196 / 49.7 & 313 / 65.0 & 313 / 64.5 & 313 / 52.1 \\
Sine + Noise & 789 / 639.0 & 316 / 65.5 & 316 / 65.2 & 316 / 53.6 \\
Switching & 5 / 399.0 & 15 / 106.4 & 18 / 230.6 & 18 / 1.6 \\
\bottomrule
\end{tabular}

    \label{tab:cardinality-aad}
    \vspace{-1em}
\end{table}
The cardinality and \gls{aad} of a dataset are good indicators of its compressibility.
Observing how transformations affect these metrics helps understand their effectiveness.
Cardinality has a direct impact on entropy and, hence, lower cardinality generally suggests better compressibility.
Meanwhile, low \gls{aad} is a relevant compressibility indicator only for methods that operate according to the smaller-numbers-fewer-bits principle.
Among the reviewed methods, this is the case for Golomb coding, DRH, and Sprintz.\par
\emph{(1) If the values of a time series are sufficiently autocorrelated (the time series is ``slowly varying''), delta coding can significantly reduce the entropy of the data.}
In the sine and sine+noise test cases, this manifests in a reduction of both \gls{aad} and cardinality.
The other two test cases exhibit much less autocorrelation and delta coding actually increases the cardinality and \gls{aad}.
\emph{(2) \gls{quars} does not change the cardinality of the data but reduces the \gls{aad}.}
\gls{quars} reorders bins of the data such that the most frequent values become the smallest.
It is particularly effective for the switching data, for which it reduces the \gls{aad} by a factor of almost 200.\par
\emph{(3) Run-length encoding is most effective on data with long runs of constants.}
This is the case for the switching test case, where \gls{rle} reduces the length of the data by more than half (see the x-axis in Fig. \ref{fig:test_cases-a}).

\begin{figure*}[ht]
    \centering
    \includegraphics[width=\textwidth]{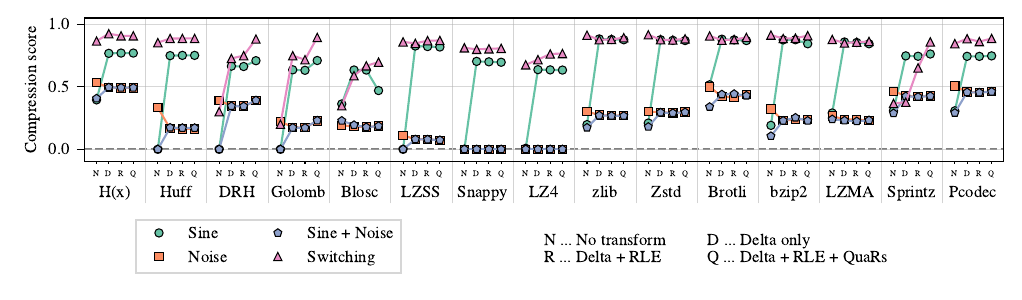}%
    \vspace{-1.5em}
    \caption{
        Compression scores achieved by the different methods on the synthetic test cases.
        The markers along a line correspond to the progressive application of delta coding, \gls{rle}, and \gls{quars}.
        Compression scores below $0$ (compressed data is larger than the original data) are shown as $0$.
    }
    \label{fig:test_cases-b}
\end{figure*}

Fig. \ref{fig:test_cases-b} shows the compression scores achieved by all methods on the test case data.
To follow the markers along lines from left to right corresponds to first adding a delta coding, then \gls{rle}, and finally a \gls{quars} transformation stage before compression.
DRH, Sprintz, and Pcodec incorporate a delta coding stage by design.
We actively disable this stage in their source code to obtain results without delta coding for illustrative purposes.\par
\emph{(4) The Shannon limit is inversely proportional to the entropy and, thus, the cardinality of the data.}
The leftmost column in Fig. \ref{fig:test_cases-b}--indicated by $H(x)$--represents the Shannon limit \cite{shannonMathematicalTheoryCommunication1948}.
This is the theoretical upper bound for the compression achievable by any method, given that the data is uncorrelated.
By exploiting correlation in the data to decrease its cardinality, delta coding can ``improve the Shannon limit'', i.e., decrease the pointwise entropy of the data.
This directly leads to better compressibility.\par
\gls{quars}, on the other hand, does not change the cardinality of the data and hence does not affect the Shannon limit.
However, it can still improve compression scores by reducing the \gls{aad}.
This can best be observed in the switching test case, where \gls{quars} improves the compression scores of Golomb, DRH, and Sprintz by 18\%, 13\%, and 21\% respectively.
All the other methods are negligibly affected by \gls{quars}, as they do not benefit from the decrease in \gls{aad}.\par
\emph{(5) Noisy signal components are a bottleneck for compression.}
If a signal is purely noise, no correlation can be exploited for better compression, rendering the Shannon limit the true upper bound for the compression achievable by any method.
In the sine+noise test case, delta coding effectively removes the low-frequency component (the sine), leaving just the noise.
Consequently, the Shannon limit (and the achieved compression scores) in the noise and sine+noise test cases are identical after delta coding.\par
\emph{(6) Overall, the hybrid and time-series-specialized methods yield the best compression scores.}
Among the general-purpose compressors, the hybrid methods are the best-performing ones.
Brotli is the best in this category.
Generally, the hybrid methods--zlib, Zstandard, Brotli, LZMA, and bzip2--outperform the dictionary-coding-only methods \gls{lzss}, Snappy, and LZ4.
This shows that deploying an entropy coding stage is crucial for achieving good compression.\par
\emph{(7) Sprintz achieves competitive compression ratios except on certain types of data.}
Decorrelating transforms such as \gls{quars} can help to mitigate these shortfalls.
\emph{(8) Pcodec is a promising allrounder method tailored to time series.}
It achieves competitive compression ratios across all sorts of time series characteristics.\par
\emph{(9) Simple static codes such as Exp-Golomb can achieve decent compression ratios on various time series, if they are paired with suitable transformations.}\par
\emph{(10) Dynamic codes such as Huffman are theoretically optimal, but the required headers quickly become a bottleneck if the data has high cardinality.}
This can be observed in the sine and sine+noise test cases, for which Huffman coding falls short of the Shannon limit due to large headers.

\subsection{Real-world time series data}
\label{subsec:real-world-time-series-data}

To obtain a performance evaluation that is closer to industrial practice and to conduct a comparison of compression speeds, we compare the compression algorithms on a selection of real-world time series datasets.
Except for ABB Drive, the datasets are open-source, inviting replication of the presented results.
They have been selected to represent a variety of domains.\par
Time series datasets are typically distributed in floating point format although in most cases, this is not how the data was originally represented (cf. \ref{subsec:time_series}).
We re-quantize the data by scaling the data to the maximum range of the chosen data type (16-bit integers) and applying the floor function.
This yields virtually no loss of precision for all datasets.
The datasets are as follows:

\subsubsection{UCR Time Series Classification Archive~\cite{UCRArchive2018}}
The UCR Archive is a collection of 128 different univariate signals.
Its intended use is as a benchmark for time series classification.

\subsubsection{PAMAP2~\cite{reissPAMAP2PhysicalActivity2012}}
A dataset featuring heart rate and IMU measurements from sensors on human subjects performing various physical activities.

\subsubsection{AMPds2~\cite{makoninAMPds2AlmanacMinutely2016}}
The Almanac of Minutely Power dataset (Version 2) contains measurements of minutely electricity, water, and natural gas consumption of a single home and over a span of two years.

\subsubsection{ABB Drive}

A dataset containing measurements from a drive system in an industrial setting.

\subsubsection{MSRC-12~\cite{fothergillInstructingPeopleTraining2012}}

The MSRC-12 Kinect Gesture Dataset contains tracking data of body-part locations.

\begin{figure*}[ht]
    \centering
    \includegraphics[width=.98\textwidth]{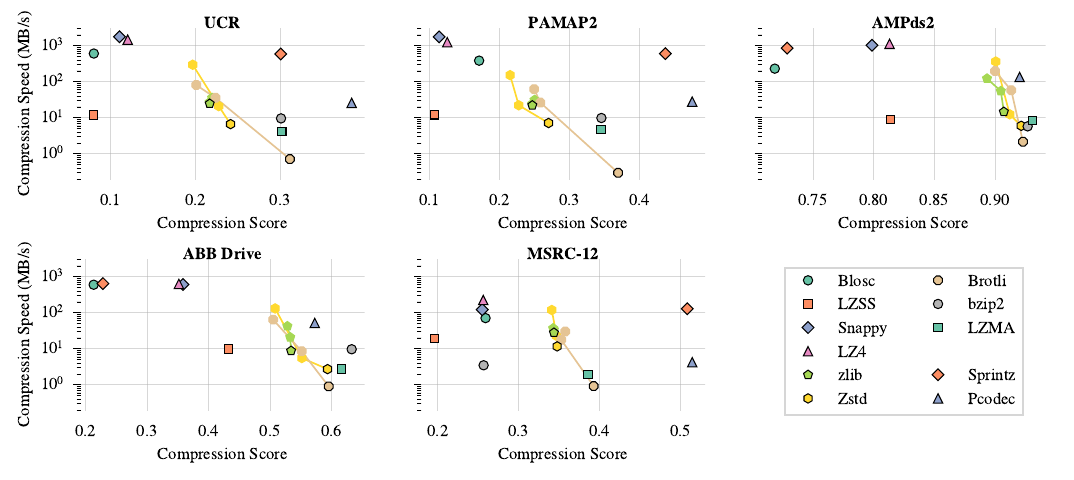}
    \vspace{-1em}
    \caption{
        Compression scores and speeds achieved by the different methods on the real-world datasets.
        Markers with no outline indicate results using compression level settings different from the default ones stated in Section \ref{subsec:experimental-setup}. 
        They indicate the speed-compression trade-off achievable by the methods Zstd, Brotli, and zlib.
    }
    \label{fig:pareto-fronts}
    \vspace{-1.3em}
\end{figure*}

Fig. \ref{fig:pareto-fronts} summarizes the compression scores and speeds achieved by the different methods on these datasets.
For these results, all methods are used as-is, i.e. Sprintz and Pcodec incorporate delta coding but no additional transformations are applied along with any method.
In contrast, Table \ref{tab:all-datasets} showcases the impact of delta coding and \gls{quars} on the compression scores for some selected methods.
We showcase the effect of \gls{quars} for Sprintz, which is one of the methods that benefit most from a small \gls{aad}.
Furthermore, we demonstrate the effect of delta coding and \gls{quars} for Brotli and bzip2, which are the best-performing general-purpose methods.
Finally, we omit any results for Huffman, Golomb coding, and DRH because their available \emph{implementations} are not competitive in terms of speed.\par

\begin{table}[t]
    \centering
    \caption{
        The impact of delta coding and \gls{quars} on the compression scores for selected methods.
    }
    \begin{tabular}{llccc}
\toprule
 &  & No transform & Delta & Delta + QuaRs \\
\midrule
UCR & Brotli & 0.31 & \textbf{0.43} & 0.41 \\
 & bzip2 & 0.30 & \textbf{0.41} & 0.39 \\
 & Sprintz &  & 0.30 & \textbf{0.34} \\
\cline{1-5}
PAMAP2 & Brotli & 0.37 & \textbf{0.48} & \textbf{0.48} \\
 & bzip2 & 0.35 & \textbf{0.47} & 0.46 \\
 & Sprintz &  & 0.44 & \textbf{0.46} \\
\cline{1-5}
AMPds2 & Brotli & \textbf{0.92} & \textbf{0.92} & \textbf{0.92} \\
 & bzip2 & \textbf{0.93} & \textbf{0.93} & \textbf{0.93} \\
 & Sprintz &  & 0.73 & \textbf{0.87} \\
\cline{1-5}
ABB Drive & Brotli & \textbf{0.60} & 0.56 & 0.56 \\
 & bzip2 & \textbf{0.63} & 0.59 & 0.58 \\
 & Sprintz &  & 0.23 & \textbf{0.45} \\
\cline{1-5}
MSRC-12 & Brotli & 0.39 & \textbf{0.51} & \textbf{0.51} \\
 & bzip2 & 0.26 & \textbf{0.37} & \textbf{0.37} \\
 & Sprintz &  & \textbf{0.51} & \textbf{0.51} \\
\bottomrule
\end{tabular}

    \label{tab:all-datasets}
    \vspace{-1em}
\end{table}

The results for the UCR, PAMAP2, and MSRC-12 datasets are qualitatively similar.
The highest compression scores achieved are around 40-55\% for these datasets.
\emph{(1) Brotli, LZMA, and bzip2 achieve good compression scores but exhibit the slowest speeds}, clustering in the lower-right portion of the graph.
Snappy and LZ4 occupy the opposite side of the graph; they compress very fast but achieve only a slight reduction of the data size.
\emph{(2) The two time-series-specialized methods Sprintz and Pcodec achieve the best trade-off of the two metrics.}
Among the two, Sprintz is the faster one, achieving speeds comparable to Snappy and LZ4 and compression scores similar to the best general-purpose methods.
Pcodec is slower than Sprintz but faster than Brotli, LZMA, and bzip2---and compresses better than any other method.
\gls{zstd}, zlib, \gls{lzss}, and Blosc lie below the Pareto front, yielding strictly worse trade-offs than other methods.\par
In Table \ref{tab:all-datasets}, one can observe the effect of the transformations on compression performance.
\emph{(3) Delta coding is a crucial component for achieving good compression scores.}
State-of-the-art general-purpose methods clearly benefit from a delta coding stage.
As can be seen in Table \ref{tab:all-datasets}, the compression scores of Brotli and bzip2 rise by more than 10\% for the UCR and PAMAP2 datasets.
This illustrates why methods such as DRH, Sprintz, and Pcodec incorporate delta coding by design.\par
For the AMPds2 and ABB Drive datasets, a few differing observations can be made.
These datasets are characterized by long runs of constants, manifests in a low cardinality but high \gls{aad}.
Consequently, they are generally more compressible than the other datasets; compression scores of up to 93\% for AMPds2 and 63\% for ABB Drive are achieved.
At the same time, their characteristics make these datasets realistic failure cases for some compression algorithms that rely on the efficient encoding of small integers.
Sprintz--being one such method--is in fact not competitive on its own on this dataset.
However, this is also a scenario in which \gls{quars} is highly effective (as discussed in \ref{subsec:test_case_suite}).
\emph{(4) After applying \gls{quars}, Sprintz achieves compression scores on par with the best-performing methods.}
Hence, combining \gls{quars} and Sprintz enables competitive compression performance at unprecedented speeds.
\section{Conclusion} \label{sec:conclusion}

We presented a comprehensive comparison of lossless time series compression algorithms using a two-stage framework of data transformation and entropy encoding.
By ablating stages and conducting the largest evaluation to date, we identified key components that drive compression performance.

Our results highlight the importance of holistic pipelines, as seen in the strong performance of hybrid compressors like bzip2, Brotli, LZMA, and specialized methods like Sprintz and Pcodec.
We further showed that applying transformations such as delta coding and QuaRs can improve compression across a variety of settings.
However, no single algorithm universally outperforms others across all datasets.

The insights from this study can guide practitioners in method selection and motivate further work toward more adaptable and efficient time series compression approaches.

\section{Future Directions} \label{sec:future}

Our findings point to several directions for future research: (1) Benchmarking efforts: Future benchmarking should evaluate full compression pipelines, emphasizing the importance of both transformation and encoding stages. High-quality, user-friendly implementations are crucial for ensuring consistent comparisons and encouraging broader adoption. Unfortunately, this is an area that still needs improvement at the time of our study.
(2) Scope expansion: Several aspects lie beyond the scope of our study, including lossy compression and streaming applications, both of which pose distinct challenges and opportunities. Future work should also consider the memory requirements and embedded-friendliness of compression methods, as well as their random access capabilities. Comprehensive comparisons along these dimensions are difficult but would be beneficial for the community.


\bibliographystyle{ieeetr}
\bibliography{ABB.bib}

\end{document}